\documentclass[11pt]{article}

\usepackage[affil-it]{authblk}
\usepackage{geometry}

\usepackage{amsmath,empheq}
\usepackage{amsfonts}
\usepackage{amsthm}
\usepackage{amssymb}
\usepackage{graphicx}
\usepackage[hyperfootnotes=false]{hyperref}
\usepackage{setspace}
\usepackage{cite}

	\newcommand{\ncd}{\newcommand}
	\ncd{\mrm}    {\mathrm}
	\ncd{\beq} {\begin{equation}}
	\ncd{\eeq} {\end{equation}}

	\def\d{{\rm d}}

\makeatletter
\def\@maketitle{%
  \newpage
  \null
  \vskip 2em%
  \begin{center}%
  \let \footnote \thanks
    {\huge \@title \par}%
    \vskip 1.5em%
    {\normalsize
      \lineskip .5em%
      \begin{tabular}[t]{c}%
        \@author
      \end{tabular}\par}%
    \vskip 1em%
    {\normalsize \@date}%
  \end{center}%
  \par
  \vskip 1.5em}
\makeatother

\makeatletter
	\renewcommand{\section}
	{\@startsection{section}{1}{0mm}
	{-\baselineskip}{0.5\baselineskip}
	{\large\bfseries}} 
\makeatother

\title{Conformal Gauge Transformations in Thermodynamics}

	\author[a]{A Bravetti%
		\thanks{ \texttt{bravetti@correo.nucleares.unam.mx}}}
		\affil[a]{Instituto de Ciencias Nucleares, Universidad Nacional Aut\'onoma de M\'exico,
		A.P. 70-543, 04510 M\'exico D.F., M\'exico.}

	\author[a]{C S Lopez-Monsalvo%
		\thanks{ \texttt{cesar.slm@correo.nucleares.unam.mx}}}

	\author[b]{F Nettel%
		\thanks{ \texttt{Francisco.Nettel@roma1.infn.it}}}
		\affil[b]{Dipartimento di Fisica, Universit\`a di Roma La Sapienza, P.le Aldo Moro 5, I-00185 Rome, Italy.}

\date{Dated: \today}

\begin{document}
\maketitle

	\begin{abstract}
	In this work we consider conformal gauge transformations of the geometric structure of thermodynamic fluctuation theory.
	In particular, we show that the Thermodynamic Phase Space is naturally endowed with a \emph{non-integrable} connection,
	defined by all those processes that annihilate the  Gibbs 1-form, i.e. reversible processes. Therefore the geometry 
	of reversible processes is invariant
	under re-scalings, that is, it has a conformal gauge freedom.
	Interestingly, as a consequence of the non-integrability of the connection, its curvature is not invariant under conformal gauge transformations and, therefore,
	neither is the associated pseudo-Riemannian geometry. We argue that this is not surprising, since these two objects are associated with irreversible
	processes.
	Moreover, 
	we  provide the explicit form in which  all the  elements of the geometric structure of the Thermodynamic Phase Space change under a conformal  gauge transformation. 
	As an example, we revisit the change of the thermodynamic representation 
	and consider the resulting change between the two metrics on the Thermodynamic Phase Space which induce Weinhold's energy metric and Ruppeiner's entropy metric. 
	As a by-product we obtain a proof of
	the well-known conformal relation between Weinhold's and Ruppeiner's metrics along the equilibrium directions. Finally,  we find 
	interesting properties of the almost para-contact structure and of its eigenvectors which may be of physical interest.
	\end{abstract}

\cleardoublepage



\section{Introduction}
The geometry of equilibrium thermodynamics and thermodynamic fluctuation theory is extremely rich.
In particular, equilibrium thermodynamics is based on the First Law, which for reversible processes can be written in the internal energy
representation as
	\beq\label{1law}
	\eta_{\rm U}=\d U-T\d S+p\,\d V-\sum_{i=1}^{n-2}\mu_{i}\d N_{i}=0,
	\eeq
where the variables have their usual meaning. From the point of view of the theory of differential equations, this is a Pfaffian system in a space of $2n+1$ variables ($n$ extensive quantities, $n$ intensities and a potential),
 for which there is no $2n$-dimensional sub-manifold whose tangent vectors all satisfy the condition \eqref{1law} (c.f. \cite{Rajeev}).
In fact, for this to be the case, the $1$-form $\eta_{\rm U}$ should satisfy the Frobenius integrability condition, $\eta_{\rm U}\wedge\d \eta_{\rm U}=0$, whereas in thermodynamics $\eta_{\rm U}$ is as far as possible from being integrable.  That
is, it satisfies
	\beq
	\eta_{\rm U}\wedge\left(\d\eta_{\rm U}\right)^{n}\neq0.
	\eeq
This implies that the solutions to eq. \eqref{1law} have at most $n$ independent variables. Therefore, thermodynamic systems are $n$-dimensional sub-manifolds
of a $(2n+1)$-dimensional \emph{phase space} which are completely defined as the graph of the `fundamental relation', i.e. a solution of \eqref{1law} expressing the dependence of the thermodynamic potential on $n$ independent variables.
As an example, for a closed thermodynamic system the fundamental relation  is usually expressed in the form $u(s,v)$, where $u$ is the molar internal energy and $s$ and $v$ are the molar entropy and volume respectively.
The equations of state for the temperature and the pressure then follow from \eqref{1law}.
  This was already realized by Gibbs and Carath\'eodory \cite{Gibbs,Cara}, 
  who started to study the geometric properties of state functions and relate them to thermodynamic properties of systems. 
In a geometric language, we can rephrase the above statements by saying that the Thermodynamic Phase Space (TPS) is a \emph{contact manifold}, and thermodynamic systems are \emph{Legendre sub-manifolds} of the TPS \cite{Hermann,mrugala1,Arnold,GTD,CRGTD}.

A Riemannian metric can be introduced on the Legendre sub-manifold representing a thermodynamic system by means of the Hessian of a thermodynamic potential. Weinhold \cite{wein1975} was the first to realize
this fact and proposed the metric defined as the Hessian of the internal energy. For example, for a closed system 
	\beq\label{gW}
	g^{W}=\frac{\partial^{2} u}{\partial s^{2}}\d s\otimes \d s+2\frac{\partial^{2} u}{\partial s \partial v}\d s\overset{\rm s}{\otimes} \d v+\frac{\partial^{2} u}{\partial v^{2}}\d v\otimes \d v,
	\eeq 
where the symbol $\overset{\rm s}{\otimes}$ denotes the symmetric tensor product (c.f. section \ref{secII}, eq. \eqref{bigG}). 
Weinhold used the inner product induced by this metric in order to recover geometrically most of the thermodynamic relations.
Later, Ruppeiner \cite{rupp1979} introduced a related metric starting from thermodynamic fluctuation theory. In fact, the Gaussian approximation
for the probability of a fluctuation \cite{LL}
	\beq\label{gaussian}
	w=w_{0}{\rm exp}\left(-\frac{\Delta T \Delta s-\Delta p \Delta v}{2T}\right)
	\eeq
depends on the Hessian of the entropy with respect to the fluctuating (extensive) variables. 
This enables one to equip the Legendre sub-manifold corresponding to a thermodynamic system with a different Hessian metric to that
of Weinhold, namely\footnote{Acutally Ruppeiner defines his metric for an open system at fixed volume, and therefore it is defined in terms of densities variables rather than molar ones. 
However, it has become common in the literature to refer to \eqref{gR} also as Ruppeiner's metric, as we do here. Moreover, notice that the original definition
of the metric by Ruppeiner has a global sign difference with the metric considered here. Of course this difference does not change any physical result, but it
is better for us to use the opposite sign convention in order to get the same conformal factor in \eqref{g1} as in \eqref{g2}.}
	\beq\label{gR}
	g^{R}=\frac{\partial^{2} s}{\partial u^{2}}\d u\otimes \d u+2\frac{\partial^{2} s}{\partial u \partial v}\d u\overset{\rm s}{\otimes} \d v+\frac{\partial^{2} s}{\partial v^{2}}\d v\otimes \d v.
	\eeq

 The two metrics are related by a \emph{conformal} re-scaling \cite{SalamonRW}
	\beq\label{g1}
	g^{R}=-\frac{1}{T}g^{W},
	\eeq
which is exactly the same re-scaling between the two $1$-forms defining the First Law in the energy and in the entropy representation, i.e.
	\beq\label{g2}
	\eta_{\rm s}=\d s-\frac{1}{T}\d u-\frac{p}{T}\d v=-\frac{1}{T}\eta_{\rm u}.
	\eeq
In this way Legendre sub-manifolds (defining thermodynamic systems undergoing reversible processes) 
are equipped naturally with two \emph{different} Riemannian structures that are related by a conformal transformation.
Notice that this fact also implies that Legendre sub-manifolds in thermodynamics are also \emph{Hessian manifolds} (see e.g. \cite{Shima,GarciaAriza}).

The study of the metrics \eqref{gW} and \eqref{gR} has been very fruitful. 
It was found in particular that the thermodynamic length corresponding to $g^{W}$ (resp. $g^{R}$) implies a lower bound on the dissipated availability 
(resp. to the entropy production) during a finite-time thermodynamic process \cite{SalamonBerryPRL} 
and that the scalar curvature of these geometries is a measure of the stability of the system, since it diverges over the critical points of continuous phase transitions with the same critical exponents 
as for the correlation volume \cite{rupp1995,rupp2010,rupp2012}.
Moreover, these geometries are related naturally to the Fisher-Rao information metric and therefore the investigation of their geometric properties 
can be extended (mutatis mutandis) to the statistical manifold \cite{BrodyRivier} and to microscopic systems, which are characterized 
by working out of equilibrium \cite{Crooks,CrooksPRE2012,CrooksPRL2012}.
As such, the intrinsic geometric perspective of Legendre sub-manifolds of the Thermodynamic Phase Space has given new physical insights on thermodynamics itself, with direct interest for applications in realistic processes, outside the realm of abstract reversible thermodynamics.
 
So far, the geometric properties of the Thermodynamic Phase Space itself have remained less investigated. Mrugala et al. \cite{MNSS1990}
proved that one can endow naturally the TPS with an \emph{indefinite} metric structure
derived from statistical mechanics which for a closed system can be defined either as
	\beq\label{GU}
	G_{\rm u}=\eta_{\rm u}\otimes\eta_{\rm u}+ \d s \overset{\rm s}{\otimes} \d T - \d v  \overset{\rm s}{\otimes}   \d p
	\eeq
or as	
	\beq\label{GS}
	G_{\rm s}=\eta_{\rm s}\otimes\eta_{\rm s}+ \d u \overset{\rm s}{\otimes} \d \left(\frac{1}{T}\right)+\d v  \overset{\rm s}{\otimes} \d \left(\frac{p}{T}\right), 
	\eeq
depending on the thermodynamic representation being considered.
These metrics reduce to Weinhold's and Ruppeiner's metrics respectively on Legendre sub-manifolds (see also \cite{Montesinos}).
 It was proved \cite{TPSSASAKI} that such structures are as perfectly well adapted to the contact structure as they can be, and that in fact, one can
introduce also a linear endomorphism in the tangent space to the TPS  so that the manifold is equipped with a very peculiar geometry,   defining a  \emph{para-Sasaskian manifold} 
\cite{Zamkovoy2009,IVZ,paraItaly}. This in turn is the odd-dimensional analogue of the well-known K\"ahler geometry \cite{KN}. Moreover, such definition implies that the TPS contains a K\"ahler manifold along the 
$2n$ directions identified with reversible processes. The important point to notice here is that the Thermodynamic Phase Space has a very rich geometric structure, with elements stemming from 
the reversible relation -- eq. \eqref{1law} --  and others arising from irreversible fluctuations, eqs. \eqref{GU} and \eqref{GS}. Furthermore, a related although at first sight slightly different geometrical approach to thermodynamic fluctuations was also recently pursued. It was shown in \cite{GCS} 
that Generalized Complex Structures, a completely new mathematical area, can be introduced in thermodynamic fluctuation theory, especially in order to consider
thermal and quantum fluctuations on the same footing, which seems to be the case in the presence of a gravitational field.

An additional physical motivation for our study comes from previous results, where it has also been proved -- by means of contact Hamiltonian dynamics -- that the lengths
computed using the metrics \eqref{GU} and \eqref{GS} 
in the Thermodynamic Phase Space 
give a measure of the entropy production along irreversible processes identified with fluctuations \cite{CONTACTHAMTD} (see also \cite{shin-itiro}).

Here, we revisit these ideas from a different point of view, namely, that of the \emph{theory of connections}. 
In this manner, we present a novel aspect of the geometric structure of thermodynamics and thermodynamic fluctuation theory. In particular,  
we study the transformations preserving the connection defined by reversible processes. 
In fact, the physical content of the First Law resides in those processes that annihilate the $1$-form $\eta_{\rm u}$ and, therefore, at the level of an equilibrium (reversible) description 
we are presented with a physical freedom of rescaling such form through multiplication
by any non-vanishing function. This operation, known as \emph{contactomorphism} \cite{libroBlair,Boyer}, 
does not change the results of equilibrium thermodynamics. 
In this sense we call such transformations \emph{conformal gauge transformations}.
One usually encounters such transformations as the change of thermodynamic representation e.g.  from the energy 
to the entropy representation [c.f. eq. \eqref{g2}]. Moreover, the connection thus defined is necessarily non-integrable, meaning that its associated curvature 
(not to be confused with the Riemannian curvature associated with the various thermodynamic metrics) 
is non-vanishing and not invariant under conformal gauge transformations.  
Hence it follows that, albeit the equilibrium thermodynamics of reversible processes 
is independent of the representation used, the description of irreversible fluctuations along such processes does change 
depending on the choice of a particular representation.

%

 We have already noticed that the associated thermodynamic metrics on the Legendre sub-manifolds re-scale as the thermodynamic 1-form $\eta$ [c.f. eqs. \eqref{g1} and \eqref{g2}]. 
 Therefore the induced thermodynamic lengths are
related but not equivalent (c.f. \cite{Schlogl}).
This is because these lengths are associated to fluctuations and irreversible processes and, therefore, they do not share such equivalence with respect to using different potentials or representations 
(for example it is well known that in non-equilibrium thermodynamics
the two problems of minimizing dissipation and maximizing work are not equivalent; see also \cite{Santoro,LiuLu,iofra} for the definition of inequivalent thermodynamic metrics
based on the Hessian of other potentials).

In this work we consider conformal gauge transformations in their full generality in the Thermodynamic Phase Space and derive the induced transformation for any  object defining its geometric structure.
We argue that these considerations can shed light over the physical significance of these geometric objects, highlighting the ones related to a reversible situation and the ones associated with irreversible evolution.
Hopefully this description will help in the identification of geometric properties of potentials that are relevant in irreversible situations.
Finally, we notice that gauge transformations in thermodynamics were also discussed in \cite{Balian} from a different perspective.

\section{The equilibrium connection}
\label{secII}

In this section we will recall some formal developments of thermodynamic geometry. 
The interested reader is referred to   \cite{TPSSASAKI} and \cite{CONTACTHAMTD} for a detailed discussion about the statistical origin of the structures presented here.

Let us consider a thermodynamic system  with $n$ degrees of freedom. As we have argued in the Introduction, the TPS -- denoted by $\mathcal{T}$ -- is the $(2n+1)$-dimensional 
ambient space of possible thermodynamic states of {any} system.

 The Laws of Thermodynamics are \emph{universal} statements (applicable to every thermodynamic system) 
 about the nature of the processes that take place when a system {evolves} from a particular thermodynamic state to another. 
 Thus we believe that such Laws are better identified in a geometric perspective with properties of the TPS.
 In order to accommodate  such Laws, it is convenient to consider the TPS to be a {differentiable} manifold. 
 This will make the evolution meaningful in terms of {vector fields} and their corresponding {integral curves}.  

 Our central point is that the First Law of Thermodynamics \eqref{1law}
 is equivalent to defining a $2n$ dimensional connection $\Gamma$ over the TPS, which we call the \emph{equilibrium connection}.
 This is a {smooth} assignment of $2n$ \emph{horizontal} directions for the tangent vectors at each point of $\mathcal{T}$. We express this schematically by
	\beq
	\label{fl}
	\{\text{First Law of Thermodynamics at $p$}\} \equiv \{\Gamma: p\in\mathcal{T} \longrightarrow \Gamma_p \subset T_p\mathcal{T}\},
	\eeq
where we use the standard notation $T_p\mathcal{T}$ for the tangent space at a given point. At first sight, such an assignment seems to be rather abstract. 
However, we will shortly see that it takes the same \emph{local} form independently of the thermodynamic system under consideration, reflecting the universality of the First Law.

Let us agree that a curve on $\mathcal{T}$ represents a possible process.  
We say that a curve joining two points in the TPS is an \emph{equilibrium (reversible) process} if its 
tangent vector lies in the horizontal subspace $\Gamma_{p}$ with respect to the First Law. 
This statement acquires a definite meaning with the aid of a connection 1-form $\eta$. Recall that a 1-form is just a \emph{linear} 
map acting on tangent vectors. In the case of the First Law, the horizontal directions of $T_p\mathcal{T}$ are given by the vectors \emph{annihilated} by  $\eta$, that is,
	\beq
	\label{horizontal}
	 X\in\Gamma_p \iff  \eta(X) = 0. 
	\eeq

From eq. \eqref{1law} we see that the above condition on $X$ is just the requirement that the corresponding process be a reversible process.
In fact, from a geometric point of view, since $\eta$ is a contact form (see Introduction), then a theorem by Darboux ensures that around each point on the TPS one 
can assign a set of local coordinates $(w,p_a,q^a)$ -- where $a$ takes values from $1$ to $n$ -- in which $\eta$
reads
	\beq\label{Darboux}
	\eta=\d w+\sum_{a=1}^{n}p_{a}\d q^{a}.
	\eeq
It can also be justified from statistical mechanical arguments (c.f. \cite{CONTACTHAMTD}) that such coordinates are the ones which enter in the equilibrium description of
the process.
These are known in the literature as \emph{Darboux coordinates}. 
For example, for a closed system as in \eqref{g2} in the molar entropy representation the coordinates $q^{a}$ are naturally associated with the extensive variables $u$ and
$v$, the $p_{a}$ are (minus) the intensities $T^{-1}$ and $p/T$ and $w$ is the molar entropy $s$.

Note that the horizontal directions in the TPS are uniquely defined by \eqref{horizontal}, and any particular thermodynamic system at equilibrium is defined to be tangent
at every point to $\Gamma_{p}$.
Therefore the definition \eqref{horizontal} encodes the universality of the First Law of Thermodynamics. 

Now, let us find a coordinate expression for the equilibrium directions around every point of the TPS. 
These are simply the tangent vectors satisfying \eqref{horizontal}. In Darboux coordinates, a direct calculation reveals that the vectors
	\beq
	\label{horbasis}
	P^{a} = \frac{\partial}{\partial p_{a}} \quad  \text{and} \quad Q_a = \frac{\partial}{\partial q^{a}} - p_{a}\frac{\partial}{\partial w},
	\eeq
generate $2n$ linearly independent horizontal directions, that is,  
	\beq
	\label{horbasis2}
	\eta(P^a) = 0 \quad \text{and} \quad \eta(Q_a) = 0,
	\eeq
for every value of $a$. Thus, every equilibrium direction around each thermodynamic state of a given system, i.e. every element of $\Gamma_p$, is a linear combination of the vectors \eqref{horbasis}.

An interesting fact is that equilibrium directions are not \emph{propagated} along equilibrium processes. To see this, 
note that the change of the $Q_b$'s along the integral curves of $P^a$ does not  vanish identically, that is, for any smooth function $f$ on $\mathcal{T}$,
	\begin{align}
	\label{nonint}
	\left[P^a,Q_b \right] (f) 	& = \left[\frac{\partial}{\partial p_a},\frac{\partial}{\partial q^b} - p_b \frac{\partial}{\partial w} \right](f)\nonumber\\
											& =- \frac{\partial}{\partial p_a} \left(p_b \frac{\partial f}{\partial w}\right) - p_b \frac{\partial}{\partial w} \left(\frac{\partial f}{\partial p_a}\right) \nonumber\\
											& = -\delta^a_{\ b} \frac{\partial f}{\partial w}  = -\delta^a_{\ b }\xi (f).
	\end{align}
Here, $\delta^{a}_{\ b}$ is a Kronecker delta and we have introduced the vector field $\xi = \partial /\partial w$, which is known in contact geometry as the \emph{Reeb vector}. 
It is straightforward to see that $\xi$ is a `purely vertical' vector at each point of the TPS in the sense of the definition \eqref{horizontal}. In fact, it is the \emph{unique} vector field satisfying 
	\beq
	\label{reeb}
	\eta(\xi) = 1 \quad \text{and} \quad \d\eta(\xi) = 0,
	\eeq
and thus can be thought of indicating the `maximally non-equilibrium' direction at each point of the TPS. 
	
Let us observe a crucial consequence of eq. \eqref{nonint}.  
Since the set \eqref{horbasis} generates $\Gamma_{p}$ at each point in the TPS, then any non-vanishing Lie-bracket of vectors in $\Gamma_p$ will be necessarily vertical.
This means that the connection $\Gamma_p$ defined by the First Law is \emph{non-integrable}\footnote{A connection is called integrable if the Lie-bracket of any pair of horizontal vector fields is horizontal \cite{KN}.
}. 
We will return to this point in the next section when we discuss its relevance on {conformal gauge invariance}. 

Now we have a basis for the tangent space $T_p\mathcal{T}$, composed by the Reeb vector $\xi$ and the horizontal basis in \eqref{horbasis}.
Notice, however, that we do not have yet a notion of orthogonality for the vector fields $\xi$, $P^a$ and $Q_a$. 
The only information available thus far is that every tangent vector to any point in the TPS can be uniquely decomposed into a vertical part and  its equilibrium (horizontal) directions, namely 
	\beq
	\label{gentan}
	X\in T_p\mathcal{T} \iff X = X_{\xi} \xi + \sum_{a=1}^n \left( X^{\rm p}_a P^a + X_{\rm q}^a Q_a\right),
	\eeq
thus the tangent space at each point of the TPS is split  into a vertical direction and $2n$ horizontal directions defined by the First Law, namely 
	\beq
	\label{split}
	T_p\mathcal{T} = V_\xi \oplus \Gamma_p.
	\eeq

In order to introduce the notion of orthogonality between the horizontal and vertical directions, one can introduce a metric structure on the TPS.
It was found by Mrugala et al. \cite{MNSS1990} (see also \cite{TPSSASAKI})
that there is a natural choice for such metric based on statistical mechanical arguments, that is
	\beq
	\label{bigG}
	G = \eta \otimes \eta - \sum_{a=1}^n \d p_a \overset{\rm s}{\otimes} \d q^a \quad \text{where} \quad \d p_a \overset{\rm s}{\otimes} \d q^a \equiv \frac{1}{2} \left[\d p_a \otimes \d q^a + \d q^a \otimes \d p_a \right]
	\eeq
Introducing a metric at this stage raises several questions about its possible significance, 
e.g. if there is a physical quantity associated to the length of a curve, the interpretation of the curvature of its  
Levi-Civita connection, Killing symmetries, etc. None of these issues will be addressed in this work. 
We will limit ourselves to use the metric as an \emph{inner product} for the tangent vectors of $T_p\mathcal{T}$ (see \cite{CONTACTHAMTD}
for a physical interpretation of the length of particular curves on the TPS corresponding to irreversible fluctuations).

 A word of warning is in place. It can be directly verified that the metric \eqref{bigG} is not positive definite, that is, there are non-zero tangent vectors whose norm vanishes identically. To see this, remember that a metric tensor is a bi-linear map (linear in its two arguments) and hence it is completely determined by its action on a set of basis vectors.  Thus, using the decomposition \eqref{split} together with the horizontal basis \eqref{horbasis}, it follows that
	\beq
	G(\xi,\xi) = 1, \quad G(P^a,Q_b) =  -\delta^a_{\ b} \quad \text{and} \quad G(\xi,P^a) = G(\xi,Q_a) = 0.
	\eeq
Interestingly, the remaining combinations vanish identically, that is
	\beq
	\label{nullvectors}
	G(Q_a,Q_a) = G(P^a,P^a) = 0.
	\eeq
There are two important things to be noticed in the above expressions. On the one hand, the metric $G$ makes the splitting of the tangent spaces \eqref{split} orthogonal. 
On the other hand, the vectors generating the horizontal basis, equation \eqref{horbasis}, form a set of \emph{null} vectors (whose norm is zero) at every point of $\mathcal{T}$. 
In general, the norm of a vector in $T_p\mathcal{T}$ [c.f. equation \eqref{gentan}] is simply given by
	\beq
	\label{bigG0}
	G(X,X) = X_{\xi}^2 - \sum_{a=1}^n X_{\rm q}^a X^{\rm p}_a,
	\eeq
and thus we can immediately see that a linear combination of null vectors is not necessarily null.

Now we want to express  the metric tensor  in \eqref{bigG} in a coordinate free manner putting into play the role of $\eta$ and $\d \eta$ as the connection 1-form and the curvature 2-form, respectively.  In terms of the geometry of contact Riemannian manifolds the result of this derivation means that the metric \eqref{bigG} is associated and compatible with the contact 1-form $\eta$ (c.f. \cite{libroBlair,Boyer}). 
Since the equilibrium connection $\Gamma_{p}$
is non-integrable,  the action of the curvature\footnote{Since the tangent space to the TPS with the equilibrium connection is a \emph{line bundle}, the curvature form $\Omega=\d \eta+\eta\wedge\eta$ coincides with $\d\eta$.
Notice also that throughout this work we are using a convention in which the wedge product is defined \emph{with} the numerical pre-factor 1/2, as in \cite{KN}, while other 
references define such product \emph{without} such pre-factor \cite{Nakahara}. 
Therefore some formulas can look different by a factor of $1/2$ with respect to other references, as e.g. in \eqref{detaUV} and
\eqref{detaandG} (for instance with respect to \cite{TPSSASAKI}). Here we choose this convention in order to make evident the relation between the second term
in the metric and the curvature of the equilibrium connection and to match with standard references in contact geometry \cite{libroBlair,Boyer}.
} of the connection 1-form \eqref{Darboux} on pairs of horizontal vectors $U,V\in \Gamma_p$,
	\beq
	\d \eta = \sum_{a=1}^n \left[\d p_a \wedge \d q^a\right] (U,V) = \frac{1}{2}\sum_{a=1}^n \left[ \d p_a(U) \d q^a(V) - \d p_a(V) \d q^a(U)\right],
	\eeq
does not necessarily vanishes. In this case, one can observe a similarity of such action with the second term in the right hand side of \eqref{bigG}. 
Let us exhibit this fact with a short calculation. Consider the coordinate expression of the two horizontal vectors $U$ and $V$, namely 
	\beq
	\label{horUV}
	U = \sum_{a=1}^n \left[ U_a^{\rm p} P^a + U^a_{\rm q} Q_a \right] \quad \text{and} \quad 	V = \sum_{a=1}^n \left[ V_a^{\rm p} P^a + V^a_{\rm q} Q_a \right].
	\eeq
Their inner product is given by
	\begin{align}
	G(U,V) 		& = \eta(U) \eta(V) - \frac{1}{2} \sum_{a=1}^n \left[\d p_a\left( U \right)\d q^a \left( V \right) + \d q^a\left( U\right) \d p_a\left( V\right) \right]\nonumber\\
					& = -\frac{1}{2} \sum_{a=1}^n \left[  U^{\rm p}_a V^a_{\rm q}  + V_a^{\rm p} U^a_{\rm q}\right],
	\end{align}
where the contribution from the first term vanishes identically since we are assuming $U,V \in \Gamma_p$. Now, a similar calculation using the exterior derivative of the connection 1-form yields
	\begin{align}
	-\d \eta(U,V) 	& = - \frac{1}{2} \sum_{a=1}^n \left[ \d p_a\left( U \right)\d q^a \left( V \right) - \d q^a\left( U\right) \d p_a\left( V\right)\right]\nonumber\\
							& = - \frac{1}{2} \sum_{a=1}^n \left[  U^{\rm p}_a V^a_{\rm q}  - V_a^{\rm p} U^a_{\rm q}\right].\label{detaUV}
	\end{align}
There is an obvious sign difference due to the fact that the metric is a symmetric tensor whereas $\d \eta$ is anti-symmetric. 
However, we can use here the same argument used in K\"ahler geometry
and introduce a linear transformation of the tangent space at each point, namely	$\Phi: T_p\mathcal{T} \longrightarrow T_p\mathcal{T}$, such that 
	\begin{align}
	-\d\eta(\Phi U,V) 	&= -\frac{1}{2} \sum_{a=1}^n \left[ \d p_a\left(\Phi U \right)\d q^a \left( V \right) - \d q^a\left(\Phi U\right) \d p_a\left( V\right)\right]\nonumber\\
									& =-\frac{1}{2} \sum_{a=1}^n  \left[  U^{\rm p}_a V^a_{\rm q}  + V_a^{\rm p} U^a_{\rm q}\right] = G(U,V).\label{detaandG}
	\end{align}
The map $\Phi$ is known in para-Sasakian geometry as the \emph{almost para-contact structure} \cite{TPSSASAKI}.
Since $\Phi$ is a linear map, it is \emph{uniquely} determined by its action on the basis vectors. Thus, one can quickly verify that the desired transformation as to satisfy
	\beq
	\label{phiaction}
	\Phi \xi = 0, \quad \Phi P^a = P^a \quad \text{and} \quad \Phi Q_a = - Q_a.
	\eeq
Thus a local expression for
	$
	\Phi: T_p\mathcal{T} \longrightarrow \Gamma_p
	$	
in this adapted basis is simply
	\beq\label{localPhi}
	\Phi=\d p_{a}\otimes P^{a}-\d q^{a}\otimes Q_{a}.
	\eeq

Now we can replace the coordinate dependent part in equation \eqref{bigG} with an equivalent purely geometric (coordinate independent) expression.
Furthermore, since $\d \eta$ `kills' the vertical part of any tangent vector [c.f. eq. \eqref{reeb}], 
our expressions are carried to any tangent vector. 
Therefore, for any pair of tangent vectors in $T_p\mathcal{T}$, their inner product is given by
	\beq
	G(X,Y) = \eta(X) \eta(Y) - \d \eta(\Phi X, Y),
	\eeq
that is, we can use a short-hand notation to  re-write equation \eqref{bigG} as
	\beq
	\label{bigG2}
	G =\eta \otimes \eta - \d \eta \circ \left(\Phi \otimes \mathbb{I} \right),
	\eeq
where $\circ$ stands for composition and $\mathbb{I}$ is the identity map on $T_p\mathcal{T}$.
	
Our final expression for the metric poses a compelling geometric structure, expressed as the sum of the First Law's connection 1-form $\eta$ and its associated \emph{field strength}  $\d \eta$, respectively. 
This was made with the aid of an intermediate quantity $\Phi$, whose role is revealed by means of its `squared' action on any vector $X \in T_p\mathcal{T}$,
	\beq
	\label{ACS}
	\Phi^2 X = \Phi \left( \Phi X\right)	= \Phi \left( \sum_{a=1}^n \left[ X_a^{\rm p} P^a - X^a_{\rm q} Q_a \right]\right) = \sum_{a=1}^n \left[ X_a^{\rm p} P^a + X^a_{\rm q} Q_a \right],
	\eeq
returning its purely horizontal part. This can be easily expressed by
	\beq
	\label{ACS2}
	\Phi^2 = \mathbb{I} - \eta \otimes \xi.
	\eeq
Finally, $\Phi$ can be independently obtained  as the covariant derivative of $\xi$ with respect to the Levi-Civita connection of $G$, closing the hard-wired geometric circuit associated to the First Law of Thermodynamics
\cite{TPSSASAKI}.

Thus far we have re-formulated the First Law as the definition of a connection whose horizontal vector fields are  
  reversible processes  [c.f. eqs. \eqref{fl} and \eqref{horizontal}].
  This sets up a suitable framework to work out the \emph{local} symmetries shared by \emph{every} thermodynamic system, 
  that is, the various points of view in which a thermodynamic  analysis can be made without changing its physical conclusions. 
  In the present case, such conclusions are restricted to the directions in which a system can evolve, and the possible interpretation 
  (not analyzed here) of the thermodynamic length of a generic process, not necessarily an equilibrium one, by means of the metric \eqref{bigG2}. 
	In the next section we will analyze an important class of such local symmetries, i.e. conformal gauge symmetries.


\section{Conformal Gauge Symmetries in Thermodynamics}

In the previous section we presented the First Law of Thermodynamics as a connection over the TPS, that is, 
the assignment of $2n$ equilibrium directions at each point of the tangent space. 
Such directions were explicitly obtained as the ones that annihilate a 1-form whose local expression is the same for every thermodynamic system. 
There is, however, a whole class of 1-forms generating exactly the same connection, each obtained from the other through multiplication
by a non-vanishing function. This is referred here as a \emph{conformal gauge freedom}.  
Thus, the central point of this section is  to present the class of transformations that one can make leaving the equilibrium connection $\Gamma$ invariant [c.f. equation \eqref{fl}], 
together with its corresponding effect on the whole \emph{intertwined} geometric structure of thermodynamic fluctuation theory, namely the para-Sasakian structure $(\mathcal{T},\eta,\xi,\Phi,G)$.

Consider the thermodynamic connection 1-form $\eta$. 
It is easy to see that any re-scaling $\eta' = \Omega \eta$ defines the same equilibrium directions at each point as the original $\eta$, that is
	\beq\label{chgauge}
	X\in\Gamma_p \iff \Omega \eta(X) = 0.
	\eeq
Here, $\Omega$ is any smooth and non-vanishing function on $\mathcal{T}$. This means we can use indistinctly $\eta$ or $\eta'$ 
to indicate the equilibrium directions at each point of $\mathcal{T}$\footnote{Usually in contact geometry $\eta$ is called the \emph{contact form}
and infinitesimal transformations generating a re-scaling of $\eta$ as in \eqref{chgauge} are known as \emph{contactomorphisms} \cite{libroBlair,Boyer}.
Here the re-scaling in \eqref{chgauge} is not necessarily derived from the action of a diffeomorphism.}. 
This, however, does change the associated metric structure. 
In particular,  for an arbitrary re-scaling, the curvature 
of the thermodynamic connection $\eta$ is \emph{not} preserved, 
as can be immediately confirmed by considering a generic pair of horizontal vectors $U,V\in\Gamma_p$ [c.f. eq. \eqref{horUV}] and making
	\beq
	\d\eta'(U,V) = \Omega \d \eta(U,V) + \frac{1}{2}\left[ \d\Omega(U)\eta(V) - \d\Omega(V)\eta(U) \right] = \Omega \d \eta(U,V).
	\eeq
Moreover, the directions annihilated by $\d \eta$ do not coincide with those of $\d \eta'$, That is, while $\d\eta(\xi) = 0$, we have
	\beq
	\d \eta'(\xi) =  \Omega \d \eta(\xi) + \left[\d \Omega \wedge \eta \right](\xi) = \frac{1}{2}\left[\d \Omega(\xi)  \eta - \d\Omega \right],
	\eeq
where in the last equality we have used the two expressions in \eqref{reeb}.
In general, the last term does not vanish and, therefore, the orthogonality of the equilibrium split of the tangent space \eqref{split}  is not trivially preserved. 
This is a consequence of the non-integrability of the 
equilibrium connection $\Gamma$. In the following lines, we will obtain the way in which the various objects introduced in the previous section change when using a different gauge.

Let us take the defining properties of the Reeb vector field, equation \eqref{reeb}, as our starting point. We need a \emph{new} vertical vector field satisfying
	\beq
	\label{newreeb}
	\eta' (\xi') = 1 \quad \text{and} \quad  \d\eta'(\xi') = 0.
	\eeq
The first condition is easily met if we define the new vertical vector field as
	\beq
	\xi' \equiv \frac{1}{\Omega} \left(\xi + \zeta \right)
	\eeq
where we have introduced an arbitrary horizontal vector field $\zeta \in \Gamma_p$ whose exact form will be determined shortly. 

The second condition in equation \eqref{newreeb} is not as trivial. A direct evaluation yields
	\begin{align}
	\label{deta1}
	\d \eta' (\xi') 	& = \Omega \d \eta(\xi') +\frac{1}{2} \left[\d \Omega (\xi')\eta - \eta(\xi') \d \Omega \right]\nonumber\\
							& = \d \eta(\zeta) + \frac{1}{2\Omega} \left[\xi(\Omega)\eta + \d \Omega(\zeta) \eta - \d \Omega \right],
	\end{align}
where we have used the fact that $\d \Omega(\xi) = \xi(\Omega)$. 
Now, we are demanding that equation \eqref{deta1} must vanish identically, that is 
	\beq\label{equals0}
	 \d \eta(\zeta) + \frac{1}{2\Omega} \left[\xi(\Omega)\eta + \d \Omega(\zeta) \eta - \d \Omega \right]=0.
	\eeq
Evaluating the above expression on $\xi$ and recalling that $\d \eta$ annihilates $\xi$,  we obtain that
	\beq\label{new}
	 \d \Omega(\zeta) = 0.
	\eeq
 Now, recalling that $\Omega$ is fixed by the change of the gauge \eqref{chgauge}, we have obtained the desired equation for $\zeta$.
Moreover, substituting \eqref{new} back into \eqref{equals0}, we obtain the expression for the derivative of the scaling factor
	\beq
	\label{domega}
	\d \Omega = 2 \Omega \d \eta(\zeta) + \xi(\Omega) \eta.
	\eeq

From these short calculations, we can infer that the auxiliary equilibrium (horizontal) vector field $\zeta$ plays a central geometric role.  
Note that in the new gauge $\eta'$, the fundamental vertical vector field $\xi'$ is tilted with respect to its unprimed counterpart, that is, it has an horizontal component. 
However, the equilibrium directions are unaltered and, therefore, are generated by the same basis vectors \eqref{horbasis}.   Thus, we write the equilibrium split at each point as
	\beq
	T_p\mathcal{T} = V_\xi \oplus \Gamma_p = V_{\xi'} \oplus \Gamma_p.
 	\eeq

Note that the expression for $\xi'$ was obtained by requiring that its geometrical properties be the same as those of $\xi$ in the new gauge [c.f. eqs. \eqref{reeb} and \eqref{newreeb}].
From the same reasoning, in analogy with \eqref{bigG2}, we require the \emph{new} metric to be given by
	\beq
	\label{newG}
	G' = \eta' \otimes \eta' - \d \eta' \circ \left(\Phi' \otimes \mathbb{I} \right).
	\eeq
The task is to find an expression for $G'$ solely in terms of unprimed objects and, just as in deriving \eqref{bigG2}, 
this reduces to obtaining an expression for the new map $\Phi'$. 
Since $\Phi'$ is just a linear transformation of each tangent space and the horizontal directions were not changed by the new gauge, 
its action on the horizontal basis must be the same as that of $\Phi$ [c.f. eq. \eqref{localPhi}].
Therefore, in order to preserve the properties \eqref{phiaction},
we only have to guarantee that its action on  $\xi'$ vanishes. 
The most general linear expression capturing these observations is $\Phi' = \Phi + \eta \otimes Z$, where the vector field $Z$ is easily determined by the requirement $\Phi'(\xi') =0$. 
Thus, {a straightforward calculation} reveals that
	\beq
	\Phi' = \Phi  - \eta \otimes \Phi(\zeta).
	\eeq
This implies that $\Phi$ and $\Phi'$ coincide on horizontal vectors, as it has to be the case. 

Consider two vector fields $X,Y \in T_p\mathcal{T}$ and their inner product in terms of the new gauge. This is expressed by the action of \eqref{newG} as
	\beq
	\label{newG2}
	G'(X,Y)		 = \Omega^2 \eta(X) \eta(Y) - \left[\Omega \d \eta(\Phi' X,Y)  +\frac{1}{2} \d\Omega(\Phi'X) \eta(Y) - \frac{1}{2}\d \Omega(Y) \eta(\Phi' X)\right].
	\eeq
We work out each individual term inside the bracket separately. Let us do this in reverse order and start with  the last term. One can immediately see that
	\beq
	\label{bra1}
	\eta(\Phi'X) = \eta\left[\Phi X - \eta(X) \Phi(\zeta)\right] = 0
	\eeq
since both  $\Phi X$ and $\Phi \zeta$ are, by construction, horizontal. Now, using the expression we obtained for the differential of the scaling factor [c.f. equation \eqref{domega}, above], combined with the action of $\Phi'$, we can re-write the next term as
	\begin{align}
	\label{bra2}
	\d\Omega(\Phi'X) \eta(Y) 	& = 2 \Omega \d \eta(\zeta, \Phi'X) + \xi(\Omega) \eta(\Phi' X)\nonumber\\
												& = - 2 \Omega \d \eta(\Phi'X,\zeta) \nonumber\\
												& = -2 \Omega \left[\d \eta(\Phi X, \zeta) - \eta(X)\d\eta(\Phi \zeta,\zeta\right] \nonumber\\
												& = -2 \Omega \left[\d \eta(\Phi X, \zeta) + \eta(X) G(\zeta,\zeta) \right].
	\end{align}
Finally, a simple expansion of the first term yields
	\beq
	\label{bra3}
	\Omega \d \eta(\Phi' X,Y) = \Omega \d \eta(\Phi X,Y) - \Omega  \eta(X) \d\eta(\Phi \zeta,Y).
	\eeq
To conclude, note that both expressions, $\d\eta(\Phi X,\zeta)$ in \eqref{bra2} and $\d\eta(\Phi \zeta,Y)$ in \eqref{bra3}, correspond to inner products involving at least one equilibrium vector. Thus we can re-write them as $G(\zeta,X)$ and $G(\zeta, Y)$, respectively. Substituting \eqref{bra1}-\eqref{bra3} back into \eqref{newG2}, adding the null term $\Omega\left[\eta(X)\eta(Y) - \eta(X)\eta(Y)\right]$ and collecting the various resulting expressions  we obtain
	\beq
	G'(X,Y) = \Omega\left[ \Omega - 1 + G(\zeta,\zeta) \right] \eta(X) \eta(Y) + \Omega \left[G(X,Y) + \eta(X) z(Y) + \eta(Y) z(X) \right],
	\eeq
where we used the shorthand $z \equiv G(\zeta)$. Hence, our final expression for the primed metric reads
	\beq
	G' = \Omega \left[G +2 \eta \overset{\rm s}{\otimes} z \right] + \Omega \left[\Omega - 1 + G(\zeta,\zeta) \right] \eta \otimes \eta. 
	\eeq
The only ambiguity left is an exact expression for $\zeta$. However, this can be easily obtained recalling once again that $\zeta\in \Gamma_p$. Thus,  using the horizontal basis \eqref{horbasis} we can write it as
	\beq
	\zeta = \sum_{a=1}^n \left[\zeta_a^{\rm p} P^a + \zeta^a_{\rm q} Q_a \right] \implies \Phi \zeta =  \sum_{a=1}^n \left[\zeta_a^{\rm p} P^a - \zeta^a_{\rm q} Q_a \right].
	\eeq
Now, a straightforward calculation reveals that 
	\beq
	G^{-1}\left[\d \eta (\zeta)\right]	 = G^{-1} \left[\sum_{a=1}^n \d p_a \wedge \d q^a \left( \sum_{b=1}^n \left[\zeta_b^{\rm p} P^b + \zeta^b_{\rm q} Q_b \right] \right)\right] = - \sum_{a=1}^n \left[ \zeta^{\rm p}_a P^a  - \zeta^a_{\rm q} Q_a\right],
	\eeq
where the inverse metric is given by 
	\beq
	\label{invmetric}
	G^{-1} = \xi \otimes \xi - 4 \sum_{a=1}^n P^a \overset{\rm s}{\otimes} Q_a.
	\eeq
 Finally, using \eqref{domega} to obtain the coordinate independent expression
	\beq
	\Phi \zeta = - G^{-1} \left[\d \eta(\zeta) \right] = -\frac{1}{2\Omega} \left[ G^{-1}(\d \Omega)  - \xi(\Omega) \xi \right],
	\eeq
and recalling the action of $\Phi^2$ [c.f. equations \eqref{ACS} and \eqref{ACS2} in the previous section], it follows that
	\beq
	\label{zetasol}
	\zeta = -\frac{1}{2 \Omega} \Phi\left[ G^{-1}(\d \Omega) \right].
	\eeq
Thus we have completely determined the new structures in terms of the old ones  and the scaling factor relating them.
Let us summarize the action of a change of gauge $(\mathcal{T},\eta,\xi,\Phi,G) \longrightarrow (\mathcal{T},\eta',\xi',\Phi',G')$, that is
	\begin{align}
		\label{gaugetransf0}
		\eta ' 		& = \Omega \eta,\\
		\label{gaugetransf1}
		\xi'  			&	=\frac{1}{\Omega}\left(\xi  + \zeta \right),\\
		\label{gaugetransf2}	
		\Phi' 		&	= \Phi +\frac{1}{2\Omega}\,\eta\otimes\left[G^{-1}(\d \Omega) - \xi (\Omega)\, \xi\right],\\
		\label{gaugetransf3}	
		G' 			&	= \Omega \left( G +2 \eta\overset{s}{\otimes} G(\zeta) \right) + \Omega \left[\Omega-1+G(\zeta,\zeta)\right]\eta\otimes\eta,
	\end{align}
where $\zeta$ is given by \eqref{zetasol}.

To close this section we shall make a few remarks on conformal gauge invariance in equilibrium thermodynamics, 
that is, the mathematical structures that are \emph{indistinguishable} along equilibrium processes when we make a change of gauge. 
Firstly, notice that the curvature of the thermodynamic connection 1-form is \emph{not} a conformally gauge invariant object, as opposed to a standard gauge theory. 
This is because the equilibrium connection $\Gamma_{p}$ is, by construction, non-integrable [c.f. equations \eqref{nonint} and \eqref{reeb}]. 
This can be interpreted physically by saying that thermodynamic fluctuations 
 are \emph{not} gauge invariant. 
Secondly, note  that in spite of the rather non-trivial  expression for the transformed metric, eq. \eqref{gaugetransf3}, its action on equilibrium vectors, say $U,V\in \Gamma_p$, is remarkably simple, that is
		\beq
		\label{conformal1}
		G'(U,V) = \Omega G(U,V).
		\eeq
Thus, in the primed gauge, the inner product  between  the basis vectors \eqref{horbasis} for the horizontal space $\Gamma_p$ is	
	\beq
	G'(P^a,Q_b) = -\Omega \delta^a_{\ b}, \quad G'(P^a,P^b) = 0 \quad \text{and} \quad G'(Q_a,Q_b) = 0.
	\eeq
Notoriously, one can immediately see that the {null equilibrium directions} at each point of the TPS  are exactly the same.  Thus, \emph{the null structure is gauge invariant}. 
Thirdly, the linear transformation $\Phi$ that we introduced on the tangent space at each point of $\mathcal{T}$ 
to obtain a coordinate free expression for the metric tensor is also a gauge invariant object with respect to equilibrium processes, 
	\beq
	\Phi' U = \Phi U \quad \text{for every} \quad U\in \Gamma_p.
	\eeq
Thus, combining the statistical origin of the metric \cite{MNSS1990,TPSSASAKI} 
and the fact that its null directions  are eigenvectors of $\Phi$, suggests that there is a physical role played by this structure. 
This will be the subject of future investigations. 
We believe that  quantities which can be directly linked to gauge invariant structures for equilibrium thermodynamics 
will be of great interest since, on the one hand, their meaning will have a universal scope (valid for every thermodynamic system) and, on the other,  
their values are independent of the thermodynamic representation one decides to use.


\section{Change of thermodynamic representation as a gauge transformation}

In the previous sections we explored some of the consequences of the geometrization of the First Law as a connection of the TPS. 
In this section we will study a particular example  and observe that the various thermodynamic representations are all related by conformal gauge transformations. 
It follows that, albeit the directions in which a state can evolve throguh an equlibrium path are independent of the thermodynamic representation,  
the fluctuations associated to the path will be different when using a different gauge.

Consider the conformal gauge transformation defined by
	\beq\label{etaprime}
	\eta' = \frac{1}{p_1} \eta = \frac{1}{p_1} \d w + \d q^1 + \sum_{a=2}^n \frac{p_a}{p_1} \d q^a,
	\eeq
where it is assumed that the Darboux neighborhood does not contain points where $p_1$ vanishes. Now, let us follow the prescription for a gauge transformation given by equations \eqref{zetasol}-\eqref{gaugetransf3}.  
We start by computing the horizontal vector field $\zeta$ in the definition of  $\xi'$.  Using \eqref{zetasol} together with the expression for the inverse metric \eqref{invmetric} and recalling the action of $\Phi$ on the horizontal basis \eqref{phiaction}, we have that
	\beq
	\label{zetaprime}
	\zeta = -\frac{1}{2} p_1 \Phi G^{-1}\left[\d \left(\frac{1}{p_1}\right) \right] = \frac{1}{p_1} Q_1.
	\eeq
Thus, the fundamental primed vertical vector field is given by
	\beq\label{xiprime}
	\xi ' = p_1 (\xi + \zeta) = p_1 \left(\xi + \frac{1}{p_1} Q_1 \right) = \frac{\partial}{\partial q^1},
	\eeq
where we have used the definition of the horizontal basis \eqref{horbasis} and the fact that in this coordinates $\xi= \partial/\partial w$. Indeed, it can by directly verified that 
	\beq
	\eta'(\xi') = \frac{1}{p_1} \d w\left(\frac{\partial }{\partial q^1}\right) + \d q^1 \left(\frac{\partial}{\partial q^1} \right) + \sum_{a=2}^n \frac{p_a}{p_1} \d q^a\left(\frac{\partial}{\partial q^1}\right) =1,
	\eeq
whereas, noting that $\partial/\partial q^1 = Q_1 + p_1 \partial/\partial w$, 
	\begin{align}
	\d \eta' \left(Q_1 + p_1 \frac{\partial}{\partial w}\right)	& = \left[ \frac{1}{p_1} \d \eta + \d \left(\frac{1}{p_1}\right) \wedge \eta \right]  \left(Q_1 + p_1 \frac{\partial}{\partial w}\right)\nonumber\\
																									& = \frac{1}{p_1} \d \eta (Q_1) - \frac{1}{2} \eta \left(Q_1 + p_1 \frac{\partial}{\partial w}\right) \d \left(\frac{1}{p_1}\right)\nonumber\\
																									& = - \frac{1}{2 p_1} \d p_1+ \frac{1}{2 p_1} \d p_1 = 0.
	\end{align}
	
 The transformation for $\Phi$ is just a straightforward calculation whose result is  
	\beq\label{Phiprime}
	\Phi ' = \Phi + \frac{p_1}{2} \eta \otimes \left[ G^{-1} \left(\d\frac{1}{p_1} \right)\right] = \Phi - \frac{1}{p_1} \eta \otimes Q_1.
	\eeq
	
Finally, in order to obtain the expression for the transformed metric,  
note that for this gauge $\zeta$ is a re-scaling of a null vector [c.f. equation \eqref{zetaprime} together with \eqref{nullvectors}]. Hence, its squared norm, $G(\zeta,\zeta)$ is identically zero. Thus, it only remains to evaluate the expression
	\beq
	G(\zeta) = -\sum_{a=1}^n \left[\d p_a \overset{\rm s}{\otimes} \d q^a\right](\zeta) = - \frac{1}{2} \sum_{a=1}^n \left[\d q^a(\zeta) \d p_a \right] = -\frac{1}{2 p_1} \d p_1. 
	\eeq
Therefore, the primed metric takes the form
	\beq\label{Gprime}
	G' = \frac{1}{p_1} \left[ G - \frac{1}{p_1} \eta \overset{\rm s}{\otimes} \d p_1 \right] + \left[\frac{1 - p_1}{p_1^2} \right]\ \eta \otimes \eta,
	\eeq
whose restriction on vectors belonging to the equilibrium connection $\Gamma_p$ at any point of the neighborhood  is simply
	\beq
	\label{conformal2}
	\left.G'\right|_{\Gamma_p} = \frac{1}{p_1} \left. G\right|_{\Gamma_p}.
	\eeq

The relevance of this exercise is that the conformal gauge transformation presented here corresponds to a change of thermodynamic representation. 
To see this, let us consider a closed system with the change of gauge defined in \eqref{g2}. 
It is clear that the equilibrium directions for both $\eta_{\rm s}$ and $\eta_{\rm u}$ are the same, as shown in the previous section.
Hence they both annihilate the vectors of the equilibrium connection $\Gamma_p$. 
Moreover, noticing that $p_{1}=-T$ in this case and that by eq. \eqref{localPhi}
	\beq
	\Phi_{\rm u}= -\left(T\,\d s-p\,\d v\right)\otimes \frac{\partial}{\partial u}-\d s\otimes \frac{\partial}{\partial s}-\d v\otimes \frac{\partial}{\partial v}+\d T\otimes \frac{\partial}{\partial T}+\d p\otimes \frac{\partial}{\partial p}\,,
	\eeq
 we can use eqs. \eqref{xiprime}, \eqref{Phiprime} and \eqref{Gprime} to obtain
	\begin{align}
	\label{gaugetransf1TD}
	\xi_{\rm u}'  & 	=-T\left[\frac{\partial}{\partial u}-\frac{1}{T}\left(\frac{\partial}{\partial s}
	+T\frac{\partial}{\partial u}\right)\right]=\frac{\partial}{\partial s}=\xi_{\rm s},\\
	\label{gaugetransf2TD}	
	\Phi_{\rm u}' &=\Phi_{\rm u}-\frac{1}{T}\eta_{\rm u}\otimes{Q}_{1}=\Phi_{\rm u}+\left(\d s-\frac{1}{T}\d u-\frac{p}{T}\d v\right)\otimes \left(\frac{\partial}{\partial s}+T\frac{\partial}{\partial u}\right)\nonumber\\
	           &= -\left(T\,\d s-p\,\d v\right)\otimes \frac{\partial}{\partial u}-\d s\otimes \frac{\partial}{\partial s}-\d v\otimes \frac{\partial}{\partial v}
	           +\d T\otimes \frac{\partial}{\partial T}+\d p\otimes \frac{\partial}{\partial p}
	           \nonumber\\
	&=
	-\left(\frac{1}{T}\d u+\frac{p}{T}\d v\right) \otimes\frac{\partial}{\partial s}-\d u\otimes \frac{\partial}{\partial u}-\d v\otimes \frac{\partial}{\partial v}
	           +\d T\otimes \frac{\partial}{\partial T}+\d p\otimes \frac{\partial}		{\partial p}=\Phi_{\rm s},\\
	\label{gaugetransf3TD}	
	G_{\rm u}' &= -\frac{1}{T} \left( G_{\rm u}+\frac{1}{T}\,\eta_{\rm u}\,\overset{s}{\otimes} \,\d T\right)+\frac{1}{T} \left(\frac{1}{T}+1\right)\eta^{U}\otimes\eta^{U}\nonumber\\
	&=-\frac{1}{T} \left( \eta_{\rm u}\otimes\eta_{\rm u}+\d s\overset{s}{\otimes}\d T-\d v\overset{s}{\otimes}\d p
			+\frac{1}{T}\,\eta_{\rm u}\,\overset{s}{\otimes} \,\d T\right)+\frac{1}{T} \left(\frac{1}{T}+1\right)\eta_{\rm u}\otimes
			\eta_{\rm u}\nonumber\\
			&=\eta_{\rm s}\otimes\eta_{\rm s}+\d u\,\overset{s}{\otimes} \,\d\left(\frac{1}{T}\right)+\d v\,\overset{s}{\otimes} \,\d\left(\frac{p}{T}\right)=G_{\rm s} \,.
	\end{align}

Equation \eqref{gaugetransf3TD} means that the metrics $G_{\rm u}$ and $G_{\rm s}$ on $\mathcal T$ are related each other by the precise 
conformal gauge transformation that corresponds to a change in the thermodynamic representation [c.f. eqs. \eqref{g1} and \eqref{g2}]. 
Moreover, it follows that on the equilibrium connection $\Gamma_{p}$ we obtain
	\beq
	\label{gaugetransf22TD}	
	\Phi_{\rm u}|_{\Gamma_{p}}=\Phi_{\rm s}|_{\Gamma_{p}} \quad \text{and} \quad 	G_{\rm u}|_{\Gamma_{p}} =-\frac{1}{T}G_{\rm s}|_{\Gamma_{p}} \,.
	\eeq
Thus, we see explicitly that the restriction of $\Phi$ to  $\Gamma_{p}$ is invariant under conformal gauge transformations,
whereas we obtain a conformal relationship between  
$G_{\rm u}$ and $G_{\rm s}$ when they
 are restricted to $\Gamma_{p}$, which exactly induces the re-scaling between Weinhold and Ruppeiner's metrics on each Legendre sub-manifold [c.f. \eqref{g1}].

\section{Closing remarks}


In thermodynamics, equilibrium (i.e. reversible) processes are defined by the First Law  \eqref{1law}. 
In this work we have given a general geometric statement of the First Law in terms of a connection on the Thermodynamic Phase Space. 
Indeed, we have shown that \eqref{1law} defines the \emph{equilibrium connection} $\Gamma_p$ [c.f. eqs. \eqref{fl} and \eqref{horizontal}]. 
Note that the connection 1-form $\eta$ defining $\Gamma_p$ is not unique. 
Indeed, any non-vanishing re-scaling $\eta'=\Omega \eta$ shares the same kernel with $\eta$ and thus defines the same equilibrium connection.
 Therefore, we call a fixing of a particular 1-form generating $\Gamma_p$ a \emph{conformal gauge choice}. 
 The name conformal is in place to denote a difference with gauge theories such as electromagnetism, where one demands  \emph{gauge invariance} 
 on the curvature of the connection, also referred as field strength. There, a choice of gauge refers to selecting a 1-form generating the same field, 
 whereas in our case, a choice of \emph{conformal} gauge refers to selecting a 1-form generating the same connection. 
 An interesting  property of the equilibrium connection is that it is always non-integrable, which means that its curvature does not vanish, independently of the choice of the conformal gauge.

To introduce a further notion of \emph{orthogonality} between the horizontal (i.e. reversible) and vertical (i.e. irreversible) directions 
with respect to the equilibrium connection $\Gamma_p$, we followed the work of Mrugala et al. \cite{MNSS1990}
and equipped  the Thermodynamic Phase Space with the indefinite metric structure \eqref{bigG}. 
One can justify such a choice by means of the statistical mechanical arguments contained in \cite{MNSS1990} and \cite{TPSSASAKI}. 
Interestingly, the null directions of such metric correspond precisely to the basis elements generating the horizontal directions \eqref{horbasis}. 
The physical significance of such directions remains to be explored and will be the subject of future work.
Here we have given a coordinate invariant formulation \eqref{bigG2} of the metric \eqref{bigG}, which highlights the role played by the connection 1-form $\eta$
 as well as by the curvature $\d\eta$ in the definition of the distance and explicitly shows that this is an associated metric in the sense of contact Riemannian geometry \cite{libroBlair,Boyer}.

The main use of presenting equilibrium thermodynamics as a connection theory relies on the notion of \emph{gauge invariance}, 
i.e. those geometric objects which are independent of the particular gauge choice. As we have argued, in the case of conformal gauge transformations, 
the curvature of the connection 1-form is not a gauge invariant object, nor is the metric. Here, we found the explicit transformations relating the various geometric objects 
defining the Thermodynamic Phase Space under a conformal gauge transformation. The explicit formulas are summarized by equations \eqref{gaugetransf0}-\eqref{gaugetransf3}. 
From these, one can observe that the null directions of the metric are gauge invariant. 
Additionally, when restricted to horizontal directions, the tensor field $\Phi$ is also gauge invariant and the metric structures are conformally related.
As an example, we have shown that the metrics \eqref{GU} and \eqref{GS} which induce Weinhold and Ruppeiner's metrics on Legendre sub-manifolds respectively are precisely related by the conformal gauge transformation that corresponds to the change in the thermodynamic representation from energy to entropy. This in turn implies that the restriction of such metrics to the equilibrium connection $\Gamma_p$ yields
 the well known conformal relation \eqref{g1}.


Finally, let us close this work with some comments on the geometry of the equilibrium connection, its conformal gauge transformations and their physical relevance in various prospect applications.  Firstly, the construction presented here exhibits the \emph{principal bundle} nature of the Thermodynamic Phase Space. That is, we readily have a $2n$-dimensional (symplectic) base manifold together with a 1-dimensional fibre isomorphic to the real line. Such construction might be suitable to make use of the theory of characteristic classes to formulate universal statements about the nature of thermodynamic processes. Secondly, from the fact that the curvature form of the connection is not preserved by a change of thermodynamic representation together with its statistical origin, one can conclude that thermodynamic fluctuations are not gauge invariant. This is interesting because thermodynamic fluctuations enter the description of irreversible processes. Therefore our results can provide new geometric insights on the different extremization problems that one encounters in non-equilibrium thermodynamics, e.g. minimizing dissipation versus maximizing work.


\section*{Acknowledgments}

AB acknowledges the A. della Riccia Foundation (Florence, Italy) for financial support. CSLM was supported by a UNAM-DGAPA Post-doctoral Fellowship. FN  acknowledges financial support from CONACYT grant No. 207934.

\end{document}